# Néel tensor torque at the ferromagnet/antiferromagnet interface

Chao-Yao Yang[1,3], Sheng-Huai Chen[1], Chih-Hsiang Tseng[1], Chang-Yang Kuo[4,5], Hsiu-Hau Lin[2]*, Chih-Huang Lai[1,6]*

[1]*Department of Materials Science and Engineering, National Tsing Hua University, Hsinchu 300044, Taiwan.*

[2]*Department of Physics, National Tsing Hua University, Hsinchu 300044, Taiwan.*

[3]*Department of Materials Science and Engineering, National Yang Ming Chiao Tung University, Hsinchu 300093, Taiwan.*

[4]*Department of Electrophysics, National Yang Ming Chiao Tung University, Hsinchu 300093, Taiwan.*

[5]*National Synchrotron Radiation Research Center, 101 Hsin-Ann Road, Hsinchu, 300092, Taiwan*

[6]*College of Semiconductor Research, National Tsing Hua University, Hsinchu 300044, Taiwan.*

\* To whom correspondence should be addressed. hsiuhau.lin@phys.nthu.edu.tw; chlai@mx.nthu.edu.tw

**Antiferromagnets (AFMs) exhibit spin arrangements with no net magnetization, positioning them as promising candidates for spintronics applications. While electrical manipulation of the single-crystal AFMs, composed of periodic spin configurations, is achieved recently, it remains a daunting challenge to characterize and to manipulate polycrystalline AFMs. Utilizing statistical analysis in data science, we demonstrate that polycrystalline AFMs can be described using a real, symmetric, positive semi-definite, rank-two tensor, which we term the 'Néel tensor'. This tensor introduces a unique spin torque, diverging from the conventional field-like and Slonczewski torques in spintronics devices. Remarkably, Néel tensors can be trained to retain a specific orientation, functioning as a form of working memory. This attribute enables zero-field spin-orbit-torque (SOT) switching in trilayer devices featuring a heavy-metal/ferromagnet/AFM structure and is also consistent with the X-ray magnetic linear dichroism measurements. Our findings uncover hidden statistical patterns in polycrystalline AFMs and establishes the presence of Néel tensor torque, highlighting its potential to drive future spintronics innovations.**

Magneto-resistive random access memory (MRAM) is a promising candidate for the next-generation nonvolatile memory (*1,2*). It manipulates magnetization for information processing and memory storage via different types of spin torques. The most common type of spin torque is the field-like torque, which is caused by the presence of an external magnetic field or the residual magnetization. However, Slonczewski predicted a different type of spin torque, called the spin-transfer torque (*3,4*), which is caused by the interaction between the magnetization and a spin current. The spin-transfer torque is more efficient than the field-like torque and can be used to write and erase MRAM bits with less power consumption. In recent years, spin-orbit torque (SOT) MRAM is proposed (*1,2*), while still under development, it has the potential to be even more efficient than spin-transfer torque MRAM.

In current designs of MRAM, antiferromagnets (AFMs) are typically used to pin the ferromagnetic (FM) layer through exchange-bias interactions (*5,6*). This is because AFMs are magnetically "neutral", meaning that their net magnetization in the bulk is zero. However, recent developments in AFM spintronics (*7-14*) offer a number of advantages over FMs, including magnetic robustness, absence of stray fields, anomalous Hall effect, and ultrafast switching dynamics (*15-17*). For example, in the CuMnAs system (*18-20*), the field-like component of spin-orbit torque (SOT) generated from an applied electric current can be used to robustly switch the Néel order of the AFM. However, most AFM materials commonly used in practical applications, such as IrMn (*21-23*), FeMn (*24,25*) and the like, are polycrystalline and it is not yet clear whether these exciting properties also exist for the polycrystalline AFM materials.

Let us re-examine spin torques. The field-like torque originates from the magnetization interaction with a vector (external field), while the Slonczewski torque (*3,4*) is due to its interaction with a tensor (spin current). Is a similar phenomenon possible at the FM/AFM interface? Yes. Beyond the well-known exchange-bias (vector) interaction, we discovered a novel tensor interaction at the FM/AFM interface, giving rise to a previously unseen type of spin torque.

In polycrystalline AFMs, including materials like IrMn and FeMn, the spins $\vec{s}_i$ (where $i = 1,2,\ldots,N_s$) exhibit spatial randomness without apparent periodicity. Pursuing this line of reasoning, we can regard the spin components $s^x$, $s^y$, $s^z$ as random variables. Utilizing the linear factor model in machine learning (*26,27*), the statistical patterns of these AFM spins can be capture by a vector $\vec{S}_R$ and a tensor $\overleftrightarrow{N}$,

$$\vec{S}_R \equiv \sum_{i=1}^{N_s} \vec{s}_i \qquad (1)$$

$$N^{\alpha\beta} \equiv \sum_{i=1}^{N_s} n_i^\alpha n_i^\beta = \left(\sum_{i=1}^{N_s} s_i^\alpha s_i^\beta\right) - S_R^\alpha S_R^\beta \qquad (2)$$

Here $\vec{n}_i = \vec{s}_i - \vec{S}_R$ represent the fluctuation from the residual spin $\vec{S}_R$ and the superscripts $\alpha, \beta = x, y, z$ denote different spin components. It is known that the residual spin $\vec{S}_R$ most locates near the interface and is directly related to the exchange bias. The rank-two tensor $\overleftrightarrow{N}$ we introduced here, termed the Néel tensor, which captures antiferromagnetic correlations within the domain, facilitates a new type of spin torque, distinct from the field-like and Slonczewski torques.

In the following, we first showcase the emergence of a novel spin torque at the FM/AFM interface during SOT switching. Then, we move on to develop the necessary theoretical framework for Néel tensor, Néel tensor interaction and Néel tensor torque. With this framework in place, we can dig into the intriguing properties of the Néel tensor, including its training and retraining through various experimental approaches.

**An unseen torque at the FM/AFM interface.** Let us start with the unexpected experimental findings in the SOT switching (*28-31*) first. **Figure 1(a)** shows the investigated trilayer structure composed of Pt/Co/IrMn, where the bottom Pt layer generate the spin current via the spin Hall effect, the middle Co layer is a perpendicular ferromagnet with (111)-crystallographic texture and the top IrMn layer is a polycrystalline AFM with the local non-collinear 3Q spin texture (*32-34*). **Figure 1(b)** illustrates the 3Q spin texture of IrMn, featuring two distinct tetrahedron configurations: spins located at the tetrahedron corners with moments directed inward (green) and outward (orange). When IrMn is grown on Co with the (111) texture, it promotes the textured IrMn (111) in the perpendicular direction as depicted in **Figure 1(c)**.

The anomalous Hall effect of the trilayer device via the $R_{xy}$-$H_z$ together with the $R_{xy}$-$H_x$ measurement is shown in **Figure 1(d)**. The exchange bias is about 1000 Oe on the $R_{xy}$-$H_z$ curve and the centrosymmetric feature on the $R_{xy}$-$H_x$ curve suggests the robust perpendicular anisotropy of Co without significant tilting along longitudinal (*x*) direction. This finding aids in eliminating the influence of the in-plane component during SOT switching in the forthcoming data analysis.

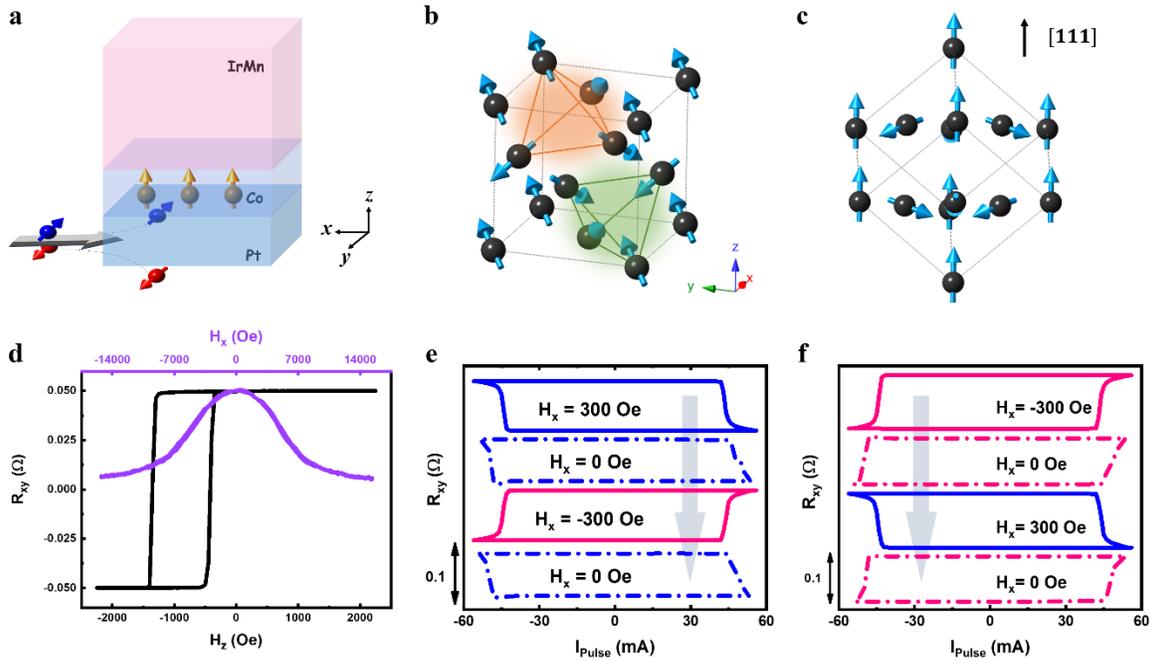

**Figure 1 | Magnetic structure of AFM IrMn and characterization on SOT switching in Pt/Co/IrMn trilayer. (a)** Schematic diagram of the SOT switching in Pt/Co/IrMn trilayer. **(b)** Magnetic structure of AFM IrMn with the 3Q spin texture. Two tetrahedrons highlighted by orange and green stand for two kinds of 3Q configurations: spins at the corner of tetrahedron pointing outward (orange) and inward (green). **(c)** Cross-sectional image of the spin texture of IrMn along [111] crystallographic direction, which is also the growth texture at the Co/IrMn interface. **(d)** The Hall resistance $R_{xy}$ of the Pt/Co/IrMn trilayer versus the perpendicular field ($H_z$) and longitudinal field ($H_x$). The exchange bias is perpendicular without any noticeable longitudinal component. **(e)** $H_x$-dependent SOT switching curves performed beginning with $H_x$=300 Oe and **(f)** $H_x$-dependent SOT switching curves performed beginning with $H_x$=-300 Oe. Arrows in the main panel of (e) and (f) represent the order of measurement in sequence.

In conventional SOT switching (*35,36*), a longitudinal magnetic field ($H_x$) is required to break the parity symmetry in the perpendicular direction through the field-like torque, as depicted in the upper part of **Figure 1(e)** (The blue solid line labeled $H_x$=300 Oe). Here comes the surprise: after completing a full SOT switching cycle, successful SOT switching is still attainable without $H_x$ (The blue dashed line labeled $H_x$=0 Oe). When an opposite field $H_x = -300$ Oe is applied, the SOT switching curve inverses its polarity as anticipated. Yet, another surprise arises: when the field is removed, the SOT switching can still be achieved but with the original polarity (The lower blue dashed line).

Note that, we carefully checked that there was no in-plane exchange bias emerged throughout each SOT switching measurement series (see Supplementary Information 1 for details). The results suggest that the device learned an intrinsic polarity once trained by a full SOT switching cycle with a magnetic field present. To confirm these unusual findings, we "trained" another as-fabricated device by a full SOT switching with an opposite field $H_x = -300$ Oe, as shown in **Figure 1(f)**. The trained device exhibits similar switching patterns, albeit with a reversed intrinsic polarity (refer to Supplementary information 2 for more details).

It is crucial to highlight that several factors dictate the SOT switching polarity, including the flowing direction of spin current ($\sigma$), sign of spin Hall angle ($\theta_{SH}$), and the direction of applied longitudinal field ($H_x$), as in the following equation:

$$P = \sigma \, \text{sign}(H_x \theta_{SH}) \qquad (3)$$

In this equation, $P = \pm 1$ stands for the switching polarity with binary values: $+1$ for counterclockwise (as shown in **Figure 1(e)** with $H_x = -300$ Oe) and $-1$ for clockwise polarity (as shown in **Figure 1(e)** with $H_x = 300$ Oe) (see Supplementary information 3 and 4 for more details). As depicted in **Figure 1(e)** and **Figure 1(f)**, the removal of $H_x$ results in an undefined polarity, unless another symmetry-breaking spin torque intervenes. Because we have carefully checked that the exchange bias does not have any in-plane component, the robust SOT switching with an intrinsic polarity implies an unseen type of spin torque at the FM/AFM interface.

**Néel tensor for polycrystalline AFM.** While the conventional Néel vector suffices for collinear AFM on bipartite lattices (*7-9*), it falls short in catching non-collinear spin configurations found in complex lattice structures like IrMn. The first-principles calculations offer a broader theoretical framework to describe AFM materials, albeit necessitating the periodicity of spin arrangements (*37-39*). The polycrystalline AFM, even though with strong local antiferromagnetic correlations, does not have a long-ranged spin order, rendering the first-principles calculations somewhat restrictive. In order to capture the antiferromagnetic correlations without long-ranged order in polycrystalline AFMs, one can resort to the linear factor model (*26,27*) utilized in stochastic data analysis within machine learning.

As explained in the previous paragraph, because the AFM spins are randomly oriented, they can be viewed as realizations of the three random variables $\vec{s} = (s^x, s^y, s^z)$. Leveraging the linear factor model, these spins in the AFM layer can be characterized by the residual spin vector $\vec{S}_R$ and the Néel tensor $\overleftrightarrow{N}$, as defined in Equations (1) and (2). It is important to note that $\vec{S}_R$ is null in an ideal antiferromagnetic domain. But, in practical situations, a non-zero $\vec{S}_R$ might emerge due to factors such as defects, distorted non-collinear spin arrangements, or uncompensated boundary spins. This residual spin plays a pivotal role in elucidating the observed exchange bias at the FM/AFM interface. Similarly, the rank-two Néel tensor $\overleftrightarrow{N}$, encapsulates the antiferromagnetic correlations between the AFM spins, portraying a covariance matrix derived from the randomness of the spin arrangements (refer to Supplementary Information 5 for more details). According to the definition, it is straightforward to show that the Néel tensor is a real, symmetric, semi-positive definite rank-two tensor and thus can be brought into diagonal form with non-negative eigenvalues with mutually orthogonal principal axes.

The interaction between the residual spin vector $\vec{S}_R$ and the magnetization vector $\vec{M}$ in a nearby FM layer manifests as the familiar exchange coupling. But what is the interaction between the Néel tensor $\overleftrightarrow{N}$ and the magnetization vector $\vec{M}$ ? Because the interaction energy is a scalar, according to tensor analysis, the simplest tensor interaction involves two "inner products" (tensor contractions in professional jargons) and takes the following form,

$$U_N = \frac{1}{2}\lambda_N \, \vec{M} \cdot \overleftrightarrow{N} \cdot \vec{M} = \frac{1}{2}\lambda_N (M_x \ M_y \ M_z) \begin{pmatrix} N^{xx} & N^{xy} & N^{xz} \\ N^{yx} & N^{yy} & N^{yz} \\ N^{zx} & N^{zy} & N^{zz} \end{pmatrix} \begin{pmatrix} M_x \\ M_y \\ M_z \end{pmatrix} \quad (4)$$

Here $\lambda_N$ denotes the coupling constant of the Néel tensor interaction $U_N$. When $\lambda_N > 0$, the Néel tensor's short axis (corresponding to the smallest eigenvalue) aligns with $\vec{M}$ to optimize the interaction energy. Conversely, for $\lambda_N < 0$, the Néel tensor's long axis (associated with the largest eigenvalue) aligns with $\vec{M}$. In the case of collinear AFMs, the Néel vector, which corresponds to the long axis of the Néel tensor, tends to orient perpendicular to the adjacent magnetization $\vec{M}$, suggesting a positive value for $\lambda_N$. In the following, we would assume $\lambda_N > 0$ in our device but the major conclusion does not rely on the sign of the coupling constant. The tensorial interaction may look formidable at first sight but similar form has been found in molecular interactions in liquid crystals (*40*), and nuclear spin interactions in nuclear magnetic resonance (*41*) with rather mature theoretical techniques.

The Néel tensor interaction can be visualized by the spherical plot, where the magnetization $\vec{M} = M\hat{r} = M(\sin\theta \cos\phi, \sin\theta \sin\phi, \cos\theta)$ is expressed in the spherical coordinates,

$$U_N = \left(\frac{1}{2}\lambda_N M^2\right)\hat{r} \cdot \overleftrightarrow{N} \cdot \hat{r} = \frac{1}{2}\lambda_N M^2 \, u_N(\theta, \phi) \quad (5)$$

The Néel tensor interaction $U_N$ is quadratic in the strength of the magnetization and its angular dependence is fully captured by the reduced Néel tensor interaction $u_N(\theta, \phi) = \hat{r} \cdot \overleftrightarrow{N} \cdot \hat{r}$, solely determined by the Néel tensor. In consequence, $u_N(\theta, \phi)$ provides the intuitive visualization of the Néel tensor interaction $U_N$, or the Néel tensor $\overleftrightarrow{N}$ itself.

Considering the current AFM-related studies, three main types of AFM texture have been reported based on spin arrangement in the space: conventional collinear AFM, non-collinear but co-planar, and neither collinear nor co-planar, as depicted in **Figure 2(a)~(c)**, respectively. The visualization of the reduced Néel tensor interaction with adjacent magnetization can refer to the polar coordinates at the bottom of **Figure 2(a)~(c)**, in which the rainbow contour reveals the strength of the interaction (red is strong and blue is weak). All spin configurations in **Figure 2** carry zero magnetization, and they can be easily characterized by the Néel tensor with spin variances along three principal axes. As elaborated in Supplementary Information 5, the long axis is defined by the maximum spin variance while the short axis is defined by the minimum. The visualization of the reduced Néel tensor thus provides a non-trivial statistical understanding of the spin configurations.

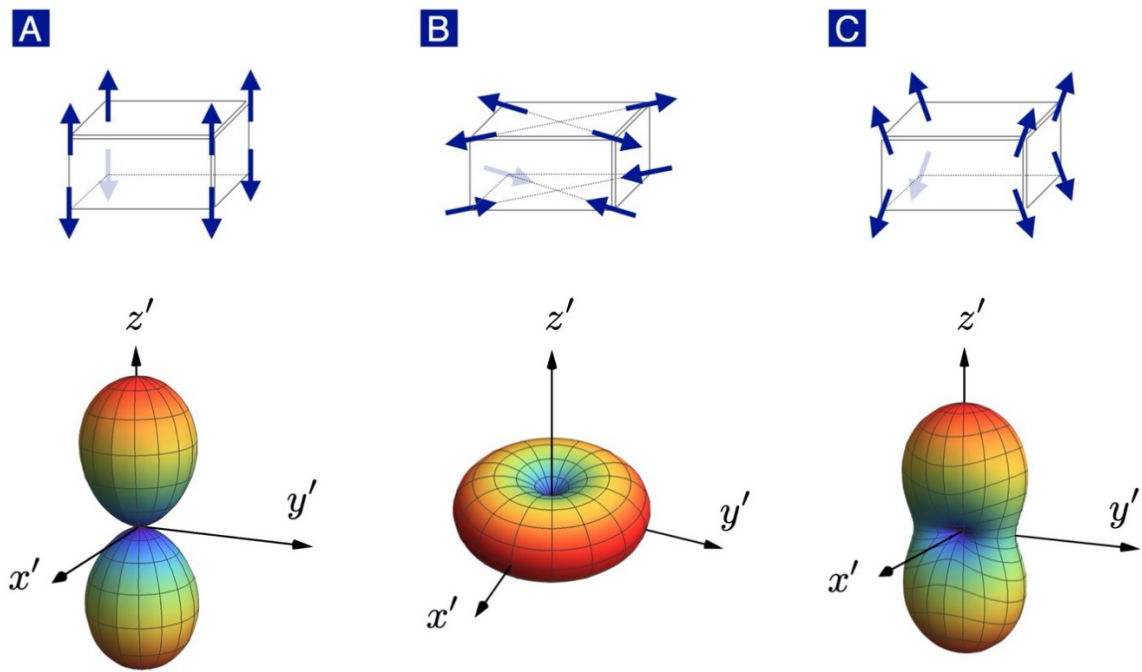

**Figure 2 | Visualizing the reduced Néel tensors.** Taking an AFM domain with eight spins as a demonstrating example, the spherical plot of the reduced Néel tensor interaction reveals the connection with the corresponding spin configurations. In the spherical plot, the value of $u_N(\theta, \phi)$ is represented as the distance to the origin. For visual aids, it is also represented by rainbow colors: the maximum value is represented by red color while the minimum value by blue color. **(a)** When all spins are collinearly aligned along the perpendicular direction, $u_N(\theta, \phi)$ reaches its maximum

(colored by red) along the $z'$-axis while its minimum (colored by blue) occurs in the transverse plane. The dumbbells-shape aligned with the $z'$-axis can be described by the Néel vector due to its axial symmetry along the long axis. **(b)** However, a non-collinear spin configuration is more challenging to comprehend. For simplicity, consider the planar spin configuration with zero component in the perpendicular direction. The maximum of $u_N(\theta, \phi)$ occurs in the $x' - y'$ plane while its minimum lies in the $z'$-axis. The spherical plot resembles the "donut" shape with degenerate maximum principal axes in the $x' - y'$ plane. **(c)** For general spin configurations with considerable non-collinearity, the reduced Néel tensor interaction resembles the "dough" shape with three orthogonal principal axes. It is clear that merely a preferential axis (represented by the Néel vector) is insufficient to capture the full feature of spin arrangements in an AFM domain.

**Néel tensor torque.** The Néel tensor interaction gives rise to a novel kind of torque exerted on the magnetization. Following the standard derivation for spin torque, an effective field is determined by computing the gradient with respect to the magnetization vector $\vec{M}$, resulting in the Néel tensor torque in the following form,

$$\vec{\tau}_N = -\vec{M} \times \nabla_M U_N = -\lambda_N M^2 \, \hat{m} \times (\overleftrightarrow{N} \cdot \hat{m}) \qquad (6)$$

Here $\hat{m}$ is the unit vector along the direction of $\vec{M}$. Note that the Néel tensor torque $\vec{\tau}_N$ encompasses both inner and outer products with magnetization and thus is qualitatively different from the Slonczewski torque $\vec{\tau}_S$ (refer to Supplementary information 6 for details).

However, although this Néel tensor may result in an additional torque to the adjacent ferromagnet, there is no symmetry-breaking effect on the SOT switching if the principal axes of Néel tensor align with the coordinate axes in the lab frame; consequently, no field-free switching can be observed, as shown in **Figure 3(a)~(c)**. On the contrary, if the principal axes of Néel tensor do not align with the coordinate axes in the lab frame, the Néel tensor torque can provide the symmetry-breaking term for the SOT switching. Consider a scenario where the short axis ($y'$ axis) of the Néel tensor is inclined in the 1st and 3rd quadrants of the $y - z$ plane (lab frame) as depicted in **Figure 3(d)**. If the magnetization points toward the positive $y$ axis, represented $\vec{M} = (0, M_y, 0)$ with $M_y > 0$, the Néel tensor torque will drive the magnetization upwards, as indicated by the arrows on the torque sphere. On the other hand, if the magnetization points towards the negative $y$ axis, represented with $\vec{M} = (0, M_y, 0)$ with $M_y < 0$, the Néel tensor torque will drive the magnetization downwards. Consequently, the orientation of the Néel tensor's short axis becomes a decisive factor in determining the magnetization switching direction, serving a symmetry-breaking function in the absence of any external field. In other words, by combining the Slonczewski and Néel tensor torques, the magnetization of the SOT device can be switched without the need for an external field, and the intrinsic polarity is determined by the alignment of the principal axes of the Néel tensor relative to the laboratory frame.

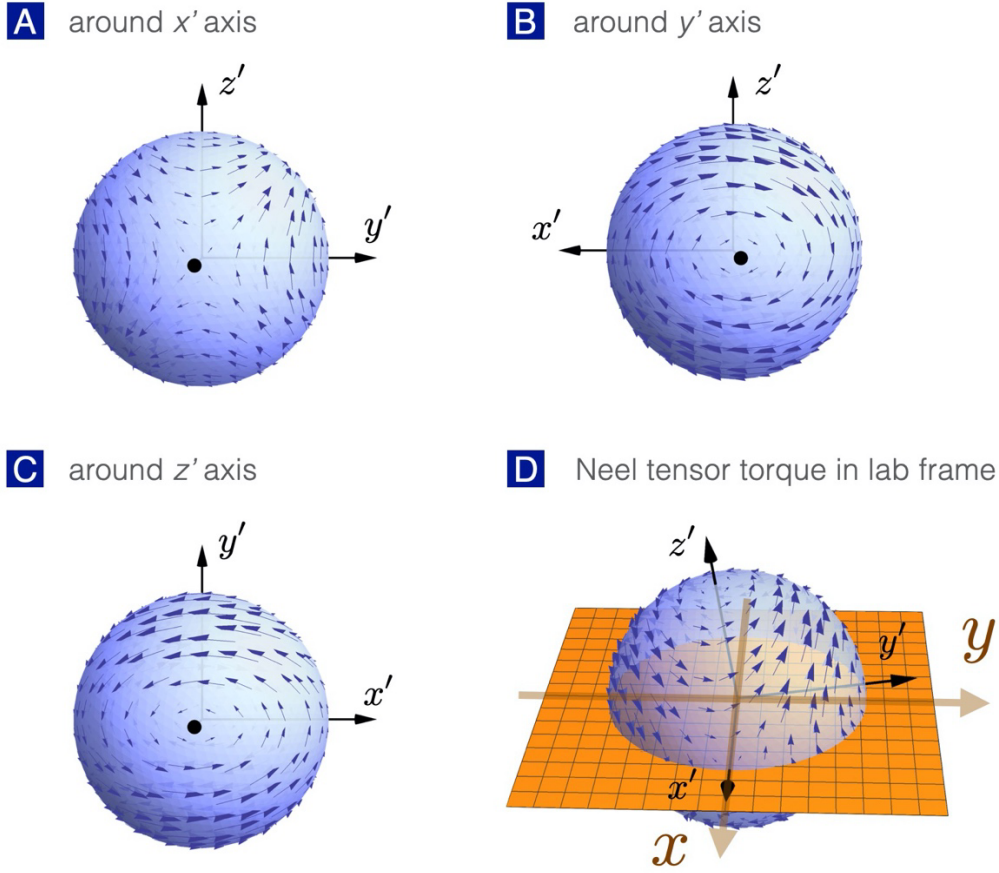

**Figure 3 | Néel tensor torque around principal axes.** The deterministic Néel tensor torque vanishes when the FM magnetization $\vec{M}$ is along one of the principal axes of Néel tensor. **(a)** The torque distributions around the $x'$ axis (intermediate spin variance) is a saddle point. **(b)** The torque distributions around the $y'$ axis (short axis) is a left-handed neutral equilibrium. **(c)** The torque distributions around the $z'$ axis (long axis) is a right-handed neutral equilibrium. **(d)** When the short axis of the Néel tensor is tilted in the 1st and the 3rd quadrants, the Néel tensor torque drives the magnetization $\vec{M} = (0, M_y, 0)$ upward or downward depending on the sign of $M_y$.

**Training the Néel tensors.** The above picture not only provides the underlying mechanism of the observed field-free SOT switching, but also leads to several interesting predictions. With longitudinal component of the exchange bias carefully ruled out, the observed SOT switching can be explained by the titled Néel tensor. Moreover, the learn-and-memorized process evokes the glassy state conceptualized in the Hopfield neural network (*42*), highlighting the crucial role of polycrystalline AFM in the functionality of the trilayer device.

To begin with, we delve into the dynamics of the Néel tensor's tilting and the reasons it maintains its orientation once established. To encapsulate the spatial inhomogeneity, we can compute the Néel tensors corresponding to each AFM domain as illustrated in **Figure 4**. Above the blocking temperature $T_B$, the AFM correlation vanishes (paramagnetic phase) and the Néel tensors are fully isotropic. The Néel tensors can thus be visualized as a collection of isotropic spheres as shown in **Figure 4(a)**. When cooled below the $T_B$, the AFM correlation molds the Néel tensors into a specific form as shown in **Figure 4(b)**. Because the principal axes of the Néel tensors for different domains are randomly oriented, there is no global order. Next, we would like to explain the effects of training process on the Néel tensors. When performing the SOT switching in the presence of the external field, the magnetization of the FM layer is tilted, as depicted in **Figure 4(c)**, and thus aligns the short axis of the Néel tensor as well. Because there are so many local minima in a polycrystalline AFM, once the Néel tensors are aligned during the training process (encompassing a full SOT switching with an external field present), they are trapped in local minima and stay put indefinitely. This is how the SOT device can learn and memorize the polarity by performing the SOT training.

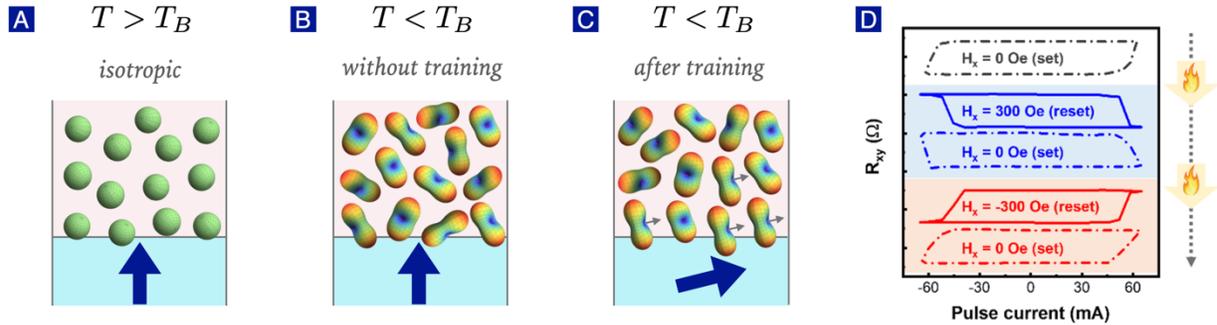

**Figure 4 | Néel glass in polycrystalline AFM.** Schematic diagrams for the Néel tensors at different temperatures. **(a)** Above the blocking temperature $T_B$, the spin arrangement is completely random and the Néel tensors become isotropic spheres. **(b)** When cooled below the blocking temperature, the Néel tensors start to take shape but the principal axes are still randomly oriented. **(c)** Going through the field-SOT training, the short axis of the Néel tensors is aligned and provides the symmetry-breaking spin torque. **(d)** Experimental verification: The trained and retrained Néel tensors are verified experimentally through the polarity inversion of the field-free SOT switching.

The memory of the SOT device can be trained and retrained as depicted in **Figure 4(d)**. Heating up the SOT device with intrinsic polarity $P = 1$ (counterclockwise) above the blocking temperature causes all Néel tensors to revert to isotropic spheres, as illustrated in **Figure 4(a)**. The memory of the SOT device is thus erased. Cooling down the device below the blocking temperature and performing the field-SOT training (with $H_x = 300$ Oe), the intrinsic polarity shifts to $P = -1$ (clockwise). Heating up the device once more, the memory is erased again.

However, cooling down the device followed by the field-SOT training with $H_x = -300$ Oe enables the device to learn and remember the polarity $P = 1$.

There are other methods to train the Néel tensors. Making use of the exchange bias in the perpendicular direction, we find an external field $\vec{H} = (0, H_y, 0)$ along the y-axis can also do the magic. As depicted in **Figure 5**, by applying a magnetic field $\vec{H} = (0, H_y, 0)$ along the y-axis, the magnetization in the FM layer is also aligned in the same direction. Together with the perpendicular exchange bias $\vec{H}_b = (0,0, H_b)$, the effective field $\vec{H}_{\text{eff}}$ lies in the $y - z$ plane and aligns the short axis of the Néel tensor accordingly. Our theory predicted that the intrinsic polarity after the training process follows the sign rule below,

$$P = \sigma \, \text{sign}(\theta_{\text{SH}}) \cdot \text{sign}(H_y H_b), \qquad (7)$$

where the derivation details can be found in the Supplementary Information 6.

The experimental results shown in **Figure 5** agrees with the theoretical predictions. In our experimental setup, the spin current is injected from the bottom layer so that $\sigma = -1$, the sign of spin Hall angle $\theta_{\text{SH}}$ (Pt layer) is positive and the perpendicular exchange bias is pointing upward. In consequence, the sign rule simplifies to $P = -\text{sign}(H_y)$. As demonstrated in **Figure 5(b)**, the training field $H_y = 7$ T leads to the intrinsic polarity $P = -1$, consistent with the sign rule's forecast. Reversing the training field to $H_y = -7$ T on the other as-deposited sample, the trained polarity shifts to $P = 1$, corroborating the prediction.

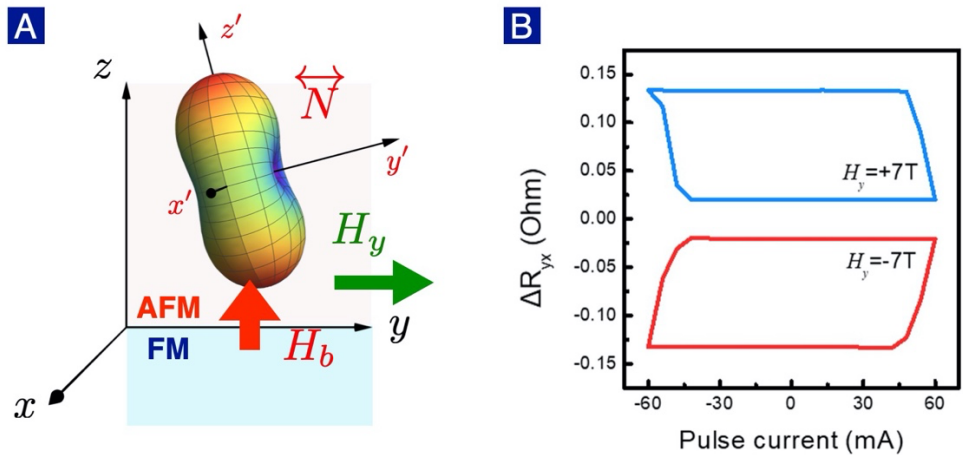

**Figure 5 | Setting Néel tensor by a magnetic field. (a)** Combining the exchange bias $\vec{H}_b = (0,0, H_b)$ and the strong field $\vec{H} = (0, H_y, 0)$, the Néel tensor can be set by the effective field $\vec{H}_{\text{eff}}$ in the $y - z$ plane. **(b)** The intrinsic polarity $P = -1$ for $H_y = 7$ T. **(c)** The intrinsic polarity $P =$

1 for $H_y = -7$ T. These results validate the predicted intrinsic polarity for the zero-field SOT switching.

**XMLD for trained Néel tensors.** X-ray magnetic linear dichroism (XMLD) provides a method to distinguish preferred AFM axes (*43,44*) within a specific plane (perpendicular to the X-ray propagation direction), establishing itself as a powerful tool for detecting tilted Néel tensors. Because the short axis of the Néel tensor after training is inclined in the $y - z$ plane, we anticipate that the X-ray absorption spectra for linear polarizations $L_x$ and $L_y$ to be different. **Figure 7(a)** and **Figure 7(b)** display the X-ray absorption spectra at the Mn $L_{2,3}$-edge, captured using two orthogonal linearly polarized X-rays (upper panel) and the corresponding XMLD (bottom panel). In this setup, we overlay the XMLD data from the SOT/field-trained device with that of a pristine device to facilitate a comparative analysis.

As a result, both SOT-trained and field-trained devices not only showcase distinguishable XMLD signals compared to the nearly noise-level signals of the pristine devices but also demonstrate a consistent dip-to-peak feature in the XMLD. This outcome is consistent with the presence of tilted Néel tensors. However, a comprehensive understanding of Néel tensor tilting requires an angular scan incorporated into the XMLD analysis.

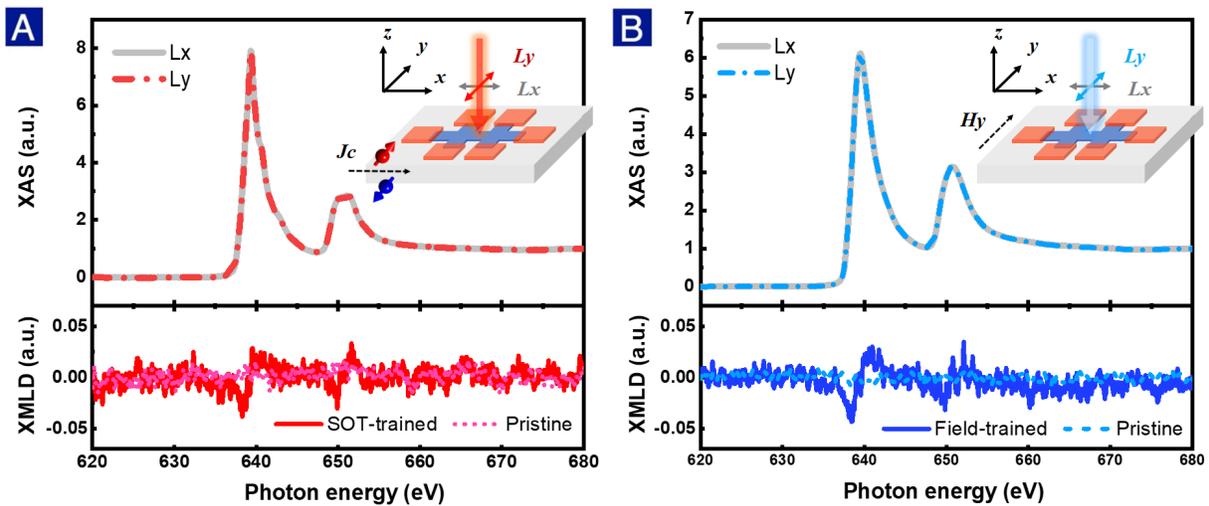

**Figure 6 | XMLD for trained Néel tensors.** In the upper panel, X-ray absorption spectra (XAS) at the Mn $L_{2,3}$-edge are illustrated for **(a)** the SOT-trained and **(b)** the field-trained devices, taken by two orthogonal linearly polarized X-rays. $L_x$ and $L_y$ refer to the linear polarization parallel and transverse to the longitudinal channel of the device. In the bottom panel, XMLD spectra, obtained from the differential of the XAS with different polarizations, are illustrated for both the trained device and the pristine device.

To comprehensively understand the Néel tensor, a 3D angular scan using XMLD analysis is imperative. While it is yet to be completed experimentally, we initiate with a theoretical exploration leveraging the configuration interaction (CI) cluster calculation (*45,46*). Within this theoretical context (see Supplementary Information 7 for details), each unique spin configuration, as determined by the CI calculation, produces predictive XMLD spectra for various X-ray incident angles. For a collinear AFM with spins oriented along the $z$ axis, the Néel vector is adequate to capture the primary characteristics. However, for non-collinear spin configurations (*47, 48*), the Néel vector alone falls short in providing a comprehensive explanation of the angular scan. This scenario points up the critical role of the Néel tensor in offering a holistic representation of spin configurations in AFM materials.

**Discussions.** While our experimental findings confirm that the Néel tensors can be trained and re-trained by different methods, the question remains: Why can the Néel tensors be trained so easily? In our polycrystalline IrMn film, its ⟨111⟩ direction is aligned perpendicularly and the AFM coupling leads to the so-called 3Q tetrahedral spin arrangements. The Néel tensor for the ideal tetrahedral spin arrangements is computed in the Supplementary Information 8 and the answer turns out to be zero.

However, in realistic materials, due to domain-domain interactions and other microscopic spin-dependent interactions, the spin arrangement in each AFM domain is not perfectly tetrahedral and the associated Néel tensor is non-vanishing below the blocking temperature. Because the energy scale of the domain-domain interaction is much weaker, it implies that the Néel tensor is more sensitive to the magnetization or external field. It is then possible to align (or modify) the Néel tensors in the polycrystalline AFM through various magnetic interactions. The success of the field training is exciting and promising because the setting process does not involve the spin current and thus can be performed effortlessly.

We have demonstrated the Néel tensor interactions with the magnetization $\vec{M}$ (and perhaps with the field $\vec{H}$), the Néel tensor can interact with the spin current $\mathcal{J}_i^\beta$ as well where the spin components are denoted by Greek indices while the spatial components by Roman indices. When injecting a spin current through a polycrystalline AFM, alterations to the spin current might occur due to its interaction with the Néel tensor. By forming the contraction between the Néel tensor $N^{\alpha\beta}$ and the spin current $\mathcal{J}_i^\beta$, it takes the following form,

$$\Delta \mathcal{J}_i^\alpha = A \sum_\beta N^{\alpha\beta} \mathcal{J}_i^\beta \qquad (8)$$

Here *A* stands for certain constant. This scenario also prompts us to consider the reverse situation: how does the spin current influence the Néel tensor? This open challenge may provide another tool to manipulate the Néel tensor and expand the horizon of AFM spintronics.

**Data availability**
The data that support the findings of this study are available from the corresponding author upon reasonable request.

**Methods**
By ultra-high vacuum magnetron sputtering system, the Pt/Co/IrMn tri-layer structures were deposited on the thermally oxidized Si substrates without external magnetic field in the following order, sub.//Ti (2nm)/Pt(2nm)/ Co(1nm)/IrMn(8nm)/Ti(5nm). The bottom and top Ti were used for the adhesion and capping layer respectively. By vibrating sample magnetometer (VSM), we confirm that all of the specimens show perpendicular magnetic anisotropy. After the examination of magnetic properties, these samples were patterned into asymmetry Hall-cross devices with 10μm and 3μm in width respectively and 40μm in length by photolithography and ion-beam etching. For all the Hall-cross devices, we further fabricated the Ta(5 nm)/Pt(100 nm) electrodes by photolithography, DC magnetron sputtering and lift-off process sequentially.

For SOT switching measurements, we took AHE signal to monitor the magnetization of the FM layer after each electric pulse. The electric pulses were applied by Keysight B2901A pulse generator with a pulse width of 0.3ms, and the AHE signal is acquired by Keithley 2000 multimeter. The out-of-plane hysteresis loops were inspected by AHE measurements.

**Acknowledgement**
This work was supported by National Science and Technology Council, Taiwan (Grant No. NSTC 111-2218-E-492-010-MBK, NSTC 111-2221-E-007-083-MY2, MOST 109-2112-M-007-026-MY3, NSTC 112-2112-M-007-035).

**Author contributions**
C.H.L. and H.H.L. planned and supervised the project. C.Y.Y carried out SOT measurements and data analysis. C.H.T and S.H.C. finished the thin film deposition and device fabrication. C.Y.K. and S.H.C. developed the configuration interaction cluster calculation. H.H.L. developed the theoretical framework and finished the calculations. H.H.L. and C.H.L. wrote the manuscript and all authors discussed the results together.

**Competing interest**

The authors declare no competing interests.

# Néel tensor torque at the ferromagnet/antiferromagnet interface

Chao-Yao Yang[1,3], Sheng-Huai Chen[1], Chih-Hsiang Tseng[1], Chang-Yang Kuo[4,5], Hsiu-Hau Lin[2]*, Chih-Huang Lai[1,6]*

[1]*Department of Materials Science and Engineering, National Tsing Hua University, Hsinchu 300044, Taiwan.*
[2]*Department of Physics, National Tsing Hua University, Hsinchu 300044, Taiwan.*
[3]*Department of Materials Science and Engineering, National Yang Ming Chiao Tung University, Hsinchu 300093, Taiwan.*
[4]*Department of Electrophysics, National Yang Ming Chiao Tung University, Hsinchu 300093, Taiwan.*
[5]*National Synchrotron Radiation Research Center, 101 Hsin-Ann Road, Hsinchu, 300092, Taiwan*
[6]*College of Semiconductor Research, National Tsing Hua University, Hsinchu 300044, Taiwan.*

* To whom correspondence should be addressed. hsiuhau.lin@phys.nthu.edu.tw; chlai@mx.nthu.edu.tw

**Methods**

Film stack of Si/Ta(2)/Pt(5)/Co(1)/IrMn(8)/Ti(5) (unit: nm) was fabricated by magnetron sputtering at room temperature with base pressure better than $2 \times 10^{-7}$ torr. Magnetic properties of sheet films were characterized by using a vibrating sample magnetometer (VSM). The sheet film structure was then patterned into the Hall-cross devices by ion beam etching with 10 μm in width and 40 μm in length, and the Ta(10)/Pt(100) electrode was deposited using an lift-off technique. Device characteristics were characterized by anomalous Hall effect (AHE) using a 4-point probe station. The AHE was acquired by applying 1 mA dc current during the collection of the voltage along the transverse channel of the Hall-cross. The SOT switching was measured via the AHE signal after giving various current amplitudes with the pulse duration of 0.3 ms. The resistance variation was performed by applying 1 mA dc current along x- and y-direction of the device before and after SOT experiments, whose signals were collected separately to avoid the electrical interference between the x- and y- channel of the Hall-cross device during the measurement.

**Supplementary Information 1 – Exclusion of the in-plane exchange bias**

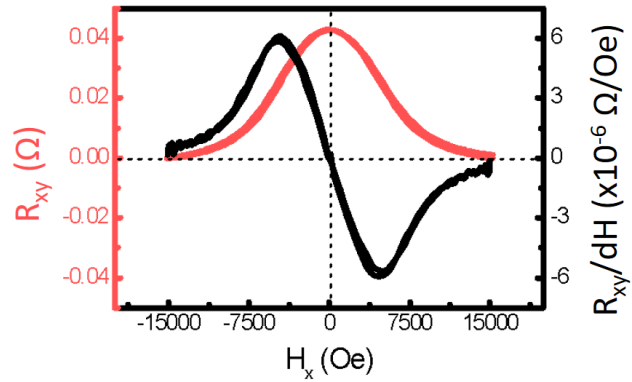

**Figure S1 | Absence of in-plane exchange bias.** Anomalous Hall effect ($R_{xy}$) versus $H_x$ curve (red) and the first-order derivative curve (black) taken after performing the SOT switching. The centro-symmetric feature on the $R_{xy}$-$H_x$ curve suggests the absence of exchange bias developed along the longitudinal direction (x-direction) of the device. The derivative $dR_{xy}/dH_x$ curve (black) is also shown across the origin to ensure the claim.

**Supplementary Information 2. – Intrinsic polarity and "switching ratio"**

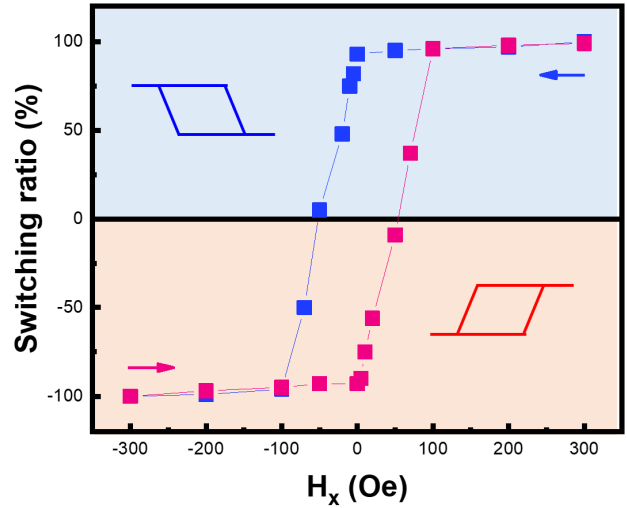

**Figure S2 | SOT switching ratio versus $H_x$.** Blue and red curves are obtained from two as-fabricated Pt/Co/IrMn devices and performed from the $H_x$-assisted switching as arrowed. The "switching ratio" is defined by the change of $R_{xy}$ on the SOT switching curve relative to that taken with $H_x = 300$ Oe. The blue (red) curve is taken by performing SOT with $H_x = 300$ Oe (-300 Oe) first, and then subsequently performing SOT with various $H_x$. Note that the polarity of blue and red curves stands for the SOT switching of $P = -1$ and $P = +1$, respectively. Two important characteristics should be noticed: the magnitudes of the switching ratio at $H_x = 0$ Oe for both cases are still larger than 90 %, suggesting the robust field-free switching as shown in **Figure 1(e)** and **Figure 1(f)** in the main text. Furthermore, once the SOT switching was performed from positive $H_x$, the switching ratio versus $H_x$ correlation is determined as a fingerprint as shown by the blue plots. The polarity of the field-free switching is always $P = -1$. After being set by $H_x = 300$ Oe (blue curve), even if we applied a small negative $H_x$ ($H_x \geq -50 \, Oe$), the polarity remains the same, indicating the Néel tensor torque prevails over the field torque under these conditions.

## Supplementary Information 3 – Sign rules from the Landau-Lifshitz-Gilbert equation

The polarity of the SOT switching is dictated by several important sign rules and can be derived from the Landau-Lifshitz-Gilbert (LLG) equation for the FM magnetization,

$$\frac{\partial \hat{m}}{\partial t} = \alpha\, \hat{m} \times \frac{\partial \hat{m}}{\partial t} - \gamma\, \hat{m} \times \vec{H}_{\text{eff}} - \gamma \sigma\, C_J\, \hat{m} \times (\hat{m} \times \hat{y}) \qquad (S1)$$

where $\hat{m} = (m_x, m_y, m_z)$ is the unit vector of the magnetization, $\alpha$ is the damping constant and $\gamma$ represents the gyromagnetic ratio. Here $\sigma = +1, -1$ denotes that the spin current is injected from top or bottom layer relative to the FM layer respectively. Two types of torques are present: the usual torque from the effective magnetic field $\vec{H}_{\text{eff}}$ and the Slonczewski torque with strength denoted by the parameter $C_J = \frac{\hbar}{2e}\frac{\theta_{\text{SH}}}{M_s t_F} J$, where $J$ is the current density, $\theta_{\text{SH}}$ is spin Hall angle, $t_F$ is the thickness of the FM layer and $M_s$ is the saturation magnetization.

The magnetization vector $\hat{m}$ quickly damps into the steady state ($\partial \hat{m}/\partial t = 0$) with direction determined by the torque balance between the longitudinal field and the spin-orbit interaction. The dynamics of the magnetization can be classified by a current density threshold,

$$J_c = \frac{e}{\hbar}\frac{M_s t_F}{\theta_{\text{SH}}} \left( H_K - \sqrt{2}\, H_x \right)$$

When the current density is below the threshold $J < J_c$, the magnetization remains in the initial direction. Above the current density threshold $J > J_c$, the torque balance gives rise to the stationary state $\hat{m}^* = (m_x^*, m_y^*, m_z^*)$,

$$m_x^* = 0 \qquad (S2)$$

$$m_y^* = \frac{1}{\sigma C_J}\sqrt{\sigma^2 C_J^2 - H_x^2} \qquad (S3)$$

$$m_z^* = \frac{H_x}{\sigma C_J} \qquad (S4)$$

The stationary magnetization $\vec{M}^* = M \hat{m}^* = M(0, m_y^*, m_z^*)$. For realistic parameters, $M_z^* \ll M_y^*$, i.e., $\vec{M}^*$ is very close to the $y$-axis with a small tilting angle $\theta_t$ away from the $x - y$ plane,

$$\theta_t = \frac{\pi}{2} - \theta = \sin^{-1}\left(\frac{H_x}{\sigma C_J}\right)$$

From the steady-state solution $\vec{M}^* = (0, M_y^*, M_z^*)$, we are now ready to derive the sign rules for the stationary magnetization during the pulse-ON period,

$$\text{sign}(M_y^*) = \sigma\, \text{sign}(C_J) = \sigma\, \text{sign}(\theta_{\text{SH}} J) \qquad (S5)$$

$$\text{sign}(M_z^*) = \sigma\, \text{sign}(H_x C_J) = \sigma\, \text{sign}(H_x \theta_{\text{SH}} J) \qquad (S6)$$

**Supplementary Information 4 – Polarity of SOT switching**

The polarity is closely related to the symmetry breaking in the SOT switching. Below the current threshold ($J < J_c$), the magnetization $\vec{M}$ in the FM layer remains in the vicinity of the $z$-axis, while it gets pushed toward the $y$-axis when the current exceeds the threshold ($J > J_c$). In the presence of longitudinal magnetic field, the Landau-Lifshitz-Gilbert (LLG) equation gives rise to a temporary stationary solution (during the pulse-on period) for the magnetization $\vec{M}^* = (0, M_y^*, M_z^*)$ where $M_z^* \ll M_y^*$, satisfying the following sign rules,

$$\text{sign}(M_y^*) = \sigma \, \text{sign}(\theta_{\text{SH}} J) \qquad (S5)$$
$$\text{sign}(M_z^*) = \sigma \, \text{sign}(H_x \theta_{\text{SH}} J) \qquad (S6)$$

Here $\sigma = +1, -1$ denotes that the spin current is injected from top or bottom layer relative to the FM layer respectively. As elaborated below, these sign rules pave the first step to understand why symmetry breaking is necessary for the SOT switching.

Define the polarity of the SOT switching as $P \equiv \text{sign}(M_z J)$. When the pulse is off, the magnetization relaxes from $M_z^*$ to the final value $M_z$ (both share the same sign) so that $P \equiv \text{sign}(M_z J) = \text{sign}(M_z^* J)$. Making use of the relation in (S5), the polarity can be rewritten in a more suggestive form,

$$P = \sigma \, \text{sign}(\theta_{\text{SH}}) \cdot \text{sign}(M_y^* M_z^*) \qquad (S7)$$

In above, we insert the trivial identity $\text{sign}(M_y^* M_y^*) = 1$ to facilitate the derivation. The first factor is related to the spin current injection, depending on the setup geometry $\sigma$ and the materials property $\theta_{\text{SH}}$. The second factor is the sign of the product $M_y^* M_z^*$, describing how parity symmetry is broken during the pulse-ON period.

Combining the sign rules $M_y^*$ and $M_z^*$ in Eq.(S5) and Eq. (S6) derived from the LLG equation, it leads to the important relation concerning symmetry breaking,

$$\text{sign}(M_y^* M_z^*) = \text{sign}(H_x) \qquad (S8)$$

It is rather remarkable that the sign of the product $M_y^* M_z^*$ solely depends on the external magnetic field $H_x$. Therefore, in the presence of the external magnetic field $H_x$, the polarity of the SOT switching is

$$P = \sigma \, \text{sign}(H_x \theta_{\text{SH}}) \qquad (S9)$$

Note that the above formula reveals the necessity of symmetry breaking for the SOT switching – the polarity $P$ is ill-defined for $H_x = 0$.

Applying the sign rule to the experimental setup described in **Figure 1**, the spin current is injected from the bottom layer so that $\sigma = -1$. The sign of spin Hall angle $\theta_{\text{SH}}$ (Pt layer) is positive and the longitudinal field $H_x$ is pointing to the positive $x$-direction. So, the polarity is $P =$ $\sigma \, \text{sign}(H_x \theta_{\text{SH}}) = -1$, consistent with the experimental finding from the $M - I$ curve. In fact, the above sign rule is verified in all our experimental results for the SOT switching.

It is important to emphasize that, after the dynamical setting, the device exhibits field-free SOT switching with an intrinsic polarity. Once the polarity is set, it remains robust and will not be erased by applying the opposite field. And, the field-free SOT switching found in our experiment is a strong hint for some unknown symmetry-breaking interaction, later identified as the Néel tensor torque.

## Supplementary Information 5 – AFM viewed as probability distribution of spin orientations

As explained in the main text, an AFM domain can be viewed as the statistical ensemble of three correlated random variables $s^x$, $s^y$, $s^z$ with specific probability distribution depending on the microscopic spin-dependent interactions. Viewing the AFM domain as the ensemble, the statistical average can be expressed in terms of summation over all spins,

$$\langle O \rangle = \frac{1}{n_d} \sum_{i=1}^{n_d} O_i \qquad (S10)$$

Here $O$ denotes an observable depending on the random variables $s^\alpha$ and $O_i$ is specific realization from the statistical ensemble. For simplicity, let us assume the residual spin $\vec{S}_R = 0$, so that the correlation matrix $\boldsymbol{Q}$, capturing the correlations between three spatial components of the spin, is simplified,

$$Q^{\alpha\beta} \equiv \langle s^\alpha s^\beta \rangle = \frac{1}{n_d} \sum_{i=1}^{n_d} s_i^\alpha s_i^\beta = \frac{1}{n_d} N^{\alpha\beta} \qquad (S11)$$

It is rather interesting to observe that the correlation matrix $\boldsymbol{Q}$ from statistical perspective is just the Néel tensor $\overleftrightarrow{N}$ defined before (up to a normalization factor $n_d$). By rotation from the lab frame to the principal frame, the correlation matrix (and also the Néel tensor) is diagonalized,

$$\boldsymbol{Q}' = \begin{pmatrix} \Delta_{x'} & 0 & 0 \\ 0 & \Delta_{y'} & 0 \\ 0 & 0 & \Delta_{z'} \end{pmatrix} \qquad (S12)$$

In the principal frame, the spin components $s^{x'}$, $s^{y'}$, $s^{z'}$ are no longer correlated because the off-diagonal elements of $\boldsymbol{Q}'$ are identically zero. The spin arrangement viewed in the principal frame is thus characterized by the variances $\Delta_{x'}$, $\Delta_{y'}$, $\Delta_{z'}$ along three principal axes. Note that "uncorrelated" is not the same as "independent" in data science and our analysis here resembles the Principal Components Analysis (PCA) within the linear factor model in machine learning and shall not be mistaken as Independent Components Analysis (ICA).

The statistical perspective provides an alternative way to look at spin arrangements in an AFM domain. Multiplying the normalization factor $n_d$, the eigenvalues of the Néel tensor are $n_d\Delta_{x'}$, $n_d\Delta_{y'}$, $n_d\Delta_{z'}$ respectively. In the principal frame, the spin components are uncorrelated and the spin variance $\Delta_{\alpha'}$ renders an intuitive picture for the spin arrangement. As shown in **Figure 2(a)**, in a collinear AFM along the $z'$-axis, spin variance $\Delta_{z'} = s^2$ (long axis) while those in the transverse directions are zero, $\Delta_{x'} = \Delta_{y'} = 0$ (short axes). In a non-collinear AFM shown in **Figure 2(c)**, the spin variances $\Delta_{z'} > \Delta_{y'} > \Delta_{x'}$, so that the spin arrangement contains largest weight along the $z'$-axis (long axis) and the smallest weight along the $x'$-axis (short axis).

**Supplementary Information 6 – Néel tensor torque**

The Néel tensor torque provides another way to understand the field-free SOT switching with intrinsic polarity. In the following, we would like to show that the intrinsic polarity satisfies the sign rule,

$$P = \sigma \, \text{sign}(\theta_{\text{SH}}) \cdot \text{sign}(n_y^* n_z^*) \quad (S13)$$

where $\vec{n}^* = (0, n_y^*, n_z^*)$ correspond to the short axis of the Néel tensor aligned in the $y$-$z$ plane. For example, if the short axis of the Néel tensor is aligned in the 1st and 3rd quadrants of the $y$-$z$ plane, as shown in **Figure 3(d)**, the intrinsic polarity is

$$P = \sigma \, \text{sign}(\theta_{\text{SH}}) \cdot \text{sign}(n_y^* n_z^*) = (-1) \cdot 1 = -1$$

The combined effects from the Slonczewski and Néel tensor torques give rise to the above sign rule. When the Slonczewski torque drives the magnetization to the positive $y$ direction, the Néel tensor torque pushes it upward and the stationary magnetization $\vec{M}^* = (0, M_y^*, M_z^*)$ lies in the 1st quadrant of the $y$-$z$ plane, as shown in **Figure 3(d)**. When the Slonczewski torque drives the magnetization to the negative $y$ direction, the Néel tensor torque pushes it downward with stationary magnetization $\vec{M}^*$ in the 3rd quadrant. In consequence, the Néel tensor torque leads to the sign rule,

$$\text{sign}(M_y^* M_z^*) = \text{sign}(n_y^* n_z^*) = 1$$

Making use of the sign rule derived in the previous paragraphs, the intrinsic polarity of the SOT switching is

$$P = \sigma \, \text{sign}(\theta_{\text{SH}}) \cdot \text{sign}(M_y^* M_z^*) = \sigma \, \text{sign}(\theta_{\text{SH}}) \cdot \text{sign}(n_y^* n_z^*) = -1$$

The sign rule reflects the short axis of the Néel tensor $\vec{n}^* = (0, n_y^*, n_z^*)$ is set by the stationary magnetization $\vec{M}^* = (0, M_y^*, M_z^*)$ during the field-SOT setting or the field setting.

**Supplementary Information 7 – Angular scan of ΔMLD**

To integrate an angular scan into the XMLD analysis, experimentally it is very challenging; therefore, we initiate with a theoretical exploration leveraging the configuration interaction (CI) cluster calculation. For simplicity, tetrahedral symmetry within the cluster is assumed and the CI calculation, encompassing thermal population with Boltzmann distribution, accurately emulates XAS at room temperature. Within this theoretical framework, each distinct spin configuration guided by the CI calculation yields predictive XMLD spectra across varying X-ray incident angles. This process facilitates the introduction of ΔMLD, derived from the differential of spectral peaks at the $L_3$-edge, serving as a reliable indicator of angular variations.

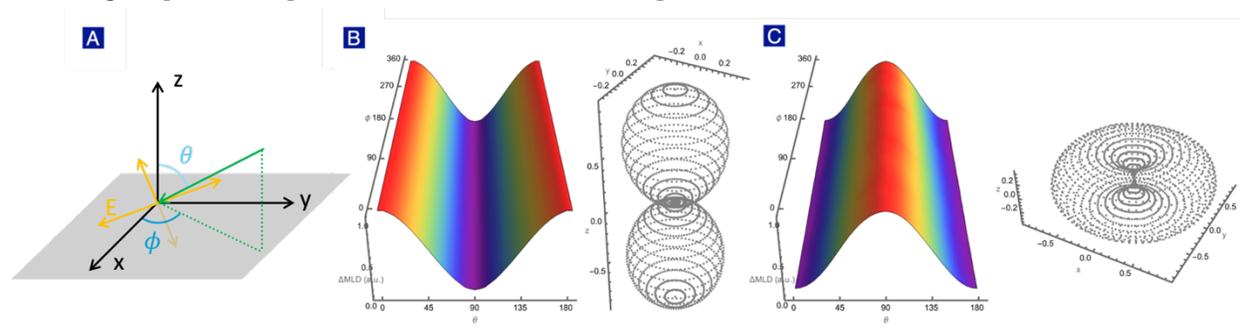

**Figure S3 | Spherical representations of ΔMLD. (a)** The X-ray propagation direction (green arrow) and corresponding electric field polarizations (yellow arrows), where $\theta$ and $\phi$ represent the angles between the $z$ and $x$ axes respectively. Since one of the electric field polarizations is lying in the $x - y$ plane during rotation, the other one can be determined by performing the outer product of the Poynting vector and the known polarization. **(b)** The angular variation of ΔMLD corresponding to the collinear spin configuration depicted in **Figure 2(a)**. The angular dependence in terms of $(\theta, \phi)$ is illustrated on the left and its corresponding spherical plot is displayed on the right. Notably, the elongated axis of the dumbbell-shaped representation aligns with the Néel vector. **(c)** This panel showcases the angular variation of ΔMLD for the non-collinear spin configuration in **Figure 2(b)**. It is evident that the conventional Néel vector is insufficient to encapsulate the donut-shaped structure emerging from the non-collinear spin arrangement.

As illustrated in **Figure S3**, we present the angular variations of ΔMLD across different spin configurations. It is important to note that we have normalized the ΔMLD values at each angle by subtracting the minimum value, facilitating a spherical plot of the angular scan. In the case of a collinear AFM where spins are aligned along the z-axis, a dumbbell shape emerges and its longest axis corresponds to the Néel vector, as depicted in **Figure S3(a)**. On the other hand, the non-collinear spin configuration leads to a donut shape as shown in **Figure S3(b)**, a formation that cannot be adequately described by the Néel vector alone. This scenario points up the critical role of the Néel tensor in offering a holistic representation of spin configurations in AFM materials.

**Supplementary Information 8 – Néel tensor for tetrahedral spin arrangement**

In our polycrystalline IrMn film, its ⟨111⟩ direction is aligned along the $z$-axis while the transverse directions are randomly oriented. The microscopic spin-dependent interaction leads to the so-called 3Q tetrahedral spin arrangement as shown in Figure **S4**. In an ideal situation, all spins at the corners point to the body center of the tetrahedron,

$$\vec{s}_1 = s(0, 0, -1)$$
$$\vec{s}_2 = s\left(-\frac{2\sqrt{2}}{3}, 0, \frac{1}{3}\right)$$
$$\vec{s}_3 = s\left(\frac{\sqrt{2}}{3}, -\frac{\sqrt{6}}{3}, \frac{1}{3}\right)$$
$$\vec{s}_4 = s\left(\frac{\sqrt{2}}{3}, \frac{\sqrt{6}}{3}, \frac{1}{3}\right)$$

The Néel tensor can be computed by summing over all spin contributions. For the ideal tetrahedral spin arrangement, the Néel tensor turns out to be zero,

$$N^{\alpha\beta} = \sum_{i=1}^{4} s_i^{\alpha} s_i^{\beta} = 0 \qquad (S14)$$

Thus, the 3Q tetrahedral spin arrangement is sensitive to all perturbations, leading to non-zero Néel tensors. However, for realistic materials below the blocking temperature, the Néel tensor associated with each domain is likely to be non-vanishing due to domain-domain interaction or/and other microscopic spin-dependent interactions, distorting the spin arrangement from the ideal 3Q tetrahedral one.

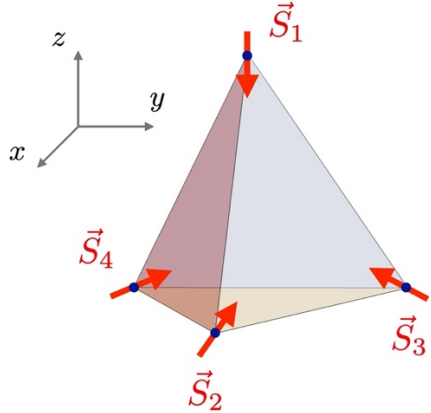

**Figure S4 | Tetrahedral spin arrangement.** In our polycrystalline IrMn film, its ⟨111⟩ direction is aligned perpendicularly and the AFM coupling leads to the so-called 3Q tetrahedral spin arrangement. It is straightforward to compute the Néel tensor for the ideal tetrahedral spin arrangement and the answer turns out to be zero.